# Chemical Precipitation Synthesis of Ferric Chloride Doped Zinc Sulphide Nanoparticles and Their Characterization Studies

T.THEIVASANTHI, N. KARTHEESWARI and M. ALAGAR

Center for Research and Post Graduate Department of Physics, Ayya Nadar Janaki Ammal College, Sivakasi – 626124, India.

*theivasanthi@pacrpoly.org*

**Abstract:** Nanoparticles of Ferric Chloride doped ZnS has been synthesized by simple chemical precipitation method and characterized by XRD, SEM, UV-Vis analysis, Differential Thermal Analysis, Thermo Gravimetric Analysis and Differential Scanning Calorimetry. XRD patterns of the samples reveal particle size, specific surface area and the formation of cubic structure. The SEM images show that the cauliflower likes structure. Optical band gap values have been obtained from UV-Vis absorption spectra. It has also been found that energy band gap ($E_g$) increases with the increase in molar concentration of reactant solution. Thermal analysis measurement of the prepared sample shows that the thermal stability of pure ZnS is decreased due to increase in Ferric Chloride concentration. Undoped ZnS is more thermal stable when compared to $FeCl_3$ doped ZnS.

## Introduction

In recent times there have been extensive studies on luminescent semiconductor nanocrystals because of their potential applications in future optoelectronic devices. The nanosized semiconductor crystallites could change optical properties which are different from bulk materials[1]. Transition metal ions and rare earth ions doped ZnS semiconductor materials have a wide range of applications in electroluminescence devices, phosphors, light emitting displays and optical sensors[2]. Semiconductor nanoparticles doped with transition metal ions have attracted wide attention due to their luminescent properties[3]. Various efforts have been made by the researchers to dope transition metal ions in nanomaterials. P.H. Borse et al have reported the luminescence quenching in ZnS nanoparticles due to Fe and Ni doping. They found the blue light emission in ZnS nanoparticles could be completely quenched when doped with iron and nickel. Generally, ZnS becomes good host material, since it is a kind of wide band gap II-VI component semiconductor materials ($E_g \sim 3.6$ eV), and with its energy band characteristics. It is commercially used as a phosphor and thin-film electroluminescence devices[4]. The characterization of the samples were carried out using X-ray diffraction (XRD), Scanning Electron Microscopy (SEM), UV-Visible absorption spectroscopy, Differential Thermal Analysis (DTA), Thermo Gravimetric Analysis (TGA) and Differential Scanning Calorimetry (DSC). From these studies, we can determine the structural, optical and thermal properties.

## Experimental Methods

Samples of different sized ZnS nanoparticles were prepared by simple chemical precipitation method using analytical grade Zinc Chloride, Ferric Chloride and Sodium Sulfide without purification. 0.1M $FeCl_3$ doped ZnS nanoparticles were prepared by mixing 0.1M of $ZnCl_2$ (20 ml) and 0.1 M of FeCl$_3$ (20 ml). Then, the obtained solution was continuously stirred for 1hours, after that 90˚C heat was supplied, along with that 0.1 M of Na$_2$S (10 ml) was slowly added to the solution and the resultant solution was continuously stirred for another 1 hours. Finally, Orange color precipitate was obtained. The obtained precipitate was allowed to evaporate at room temperature to obtain FeCl$_3$ doped ZnS nanoparticles in orange color powder form. Similarly, the preparation of 0.01M FeCl$_3$ doped ZnS nanoparticles brown color powder form is obtained.

**Results and discussions**

*XRD measurement*

The XRD Pattern of all the samples, Undoped ZnS, 0.01M FeCl$_3$ doped ZnS and 0.1M FeCl$_3$ doped ZnS is shown in Figure 1. The particle sizes, FWHM and diffraction angle values are in Table 1. As expected, the XRD peaks of nanoparticles are considerably broadened due to finite size of these particles. The prominent broad peak value of Undoped ZnS system is occurred at 28.58˚, 47.71˚ and 56.56˚and the broad peak value of 0.1 M FeCl$_3$ doped ZnS system is also occurring at 28.54˚, 45.52˚, 47.96˚ and 56.41˚. The observed diffraction peaks correspond to the (111), (220), (311) planes of the cubic crystalline ZnS are reported as identifying peaks of ZnS by earlier workers[5]. They were also compared with JCPDS file No.01-0792 and no any crystalline impurities were detected.

**Table 1.** XRD Data of Undoped ZnS and FeCl$_3$ doped ZnS

| Sample | 2θ of intense peak (deg) | FWHM β | Particle size D in (nm) | d-Spacing (Å) | hkl value | Lattice Constant a in Å |
|---|---|---|---|---|---|---|
| undoped ZnS | 28.5831 | 0.06199 | 2.41 | 3.12043 | 111 | 5.41172 |
|  | 47.7092 | 0.06979 | 2.33 | 1.90471 | 220 | 5.36089 |
|  | 56.5634 | 0.0678 | 2.43 | 1.62577 | 311 | 5.40537 |
| 0.01M FeCl$_3$ doped ZnS | 22.788 | 0.06347 | 2.33 | 3.89901 | 211 | 5.5140 |
| 0.1M FeCl$_3$ doped ZnS | 28.5455 | 0.0698 | 2.14 | 3.12446 | 111 | 5.41172 |
|  | 45.5150 | 0.0698 | 2.25 | 1.99130 | 200 | 5.63224 |
|  | 47.9594 | 0.0698 | 2.27 | 1.89536 | 220 | 5.36089 |
|  | 56.4115 | 0.0698 | 2.36 | 1.62978 | 311 | 5.40537 |

*XRD - Particle Size Calculation*

From this study, considering the peak at degrees, particle size has been estimated by using *Debye-Scherrer formula* and found to be in the range of 2.14-2.43 nm[6, 7].

$$D = \frac{0.9\lambda}{\beta cos\theta} \quad \text{............................................... (1)}$$

Where 'λ' is wave length of X-Ray (0.1541 nm), 'β' is FWHM (full width at half maximum), 'θ' is the diffraction angle and 'D' is particle diameter size. The calculated particle size details are in Table.1. The value of d (interplanar spacing between the atoms) is calculated using *Bragg's Law* [8].

$$2dsin\theta = n\lambda \quad \text{............................................... (2)}$$

Due to size effect the peaks broaden and then widths become larger as the particle size becomes smaller. The broadening of the peak may also occur due to micro strains of the crystal structure arising from defects like dislocation and twinning[9]. The standard XRD peak of ZnS is 30° and standard lattice constant is 5.41 Å[10]. The XRD shows that ZnS nanoparticle has Cubic Phase.

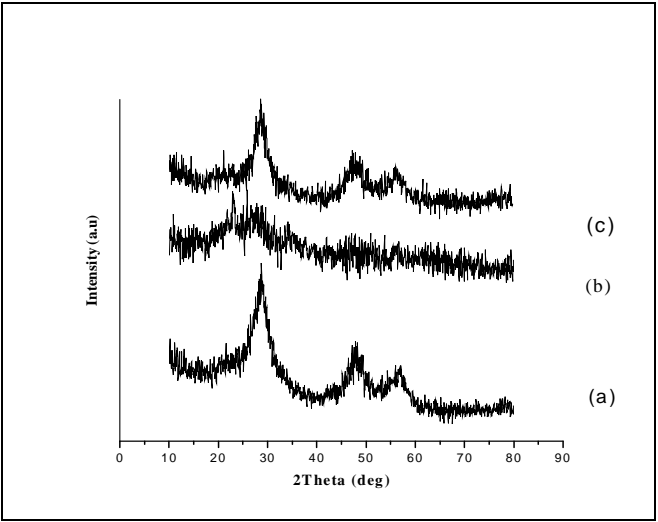

**Figure 1.** XRD Patterns of
(a) Pure ZnS  (b) 0.01 M of FeCl$_3$ doped ZnS  (c) 0.1 M of FeCl$_3$ doped ZnS

*XRD - Instrumental Broadening*

When particle size is less than 100 nm, appreciable broadening in x-ray diffraction lines will occur. Diffraction pattern will show broadening because of particle size and strain. The observed line broadening will be used to estimate the average size of the particles. The total broadening of the diffraction peak is due to the sample and the instrument. The sample broadening is described by

$$FW(S) * cos\theta = \frac{K * \lambda}{Size} + 4 * Strain * sin\theta \quad \ldots (3)$$

The total broadening $\beta_t$ is given by the equation

$$\beta_t^2 \approx \{\frac{0.9\lambda}{Dcos\theta}\}^2 + \{4\varepsilon tan\theta\}^2 + \beta_o^2 \quad \ldots (4)$$

$\varepsilon$ is strain and $\beta_0$ instrumental broadening. The average particle size D and the strain $\varepsilon$ of the experimentally observed broadening of several peaks will be computed simultaneously using *least squares method*. Instrumental Broadening of undoped ZnS and 0.1M FeCl$_3$ doped ZnS are presented in Figure 2 (a) and 2(b) respectively.

Williamson and Hall proposed a method for deconvoluting size and strain broadening by looking at the peak width as a function of 2θ. Here, Williamson-Hall plot is plotted with sin θ on the x-axis and β cos θ on the y-axis (in radians). A linear fit is got for the data. From this fit, particle size and strain are extracted from y-intercept and slope respectively[11]. The extracted particle size and strain details are enumerated in Table 2. Figure 3 (a) shows Williamson Hall Plot of undoped ZnS and 3(b) shows 0.1M FeCl$_3$ doped ZnS.

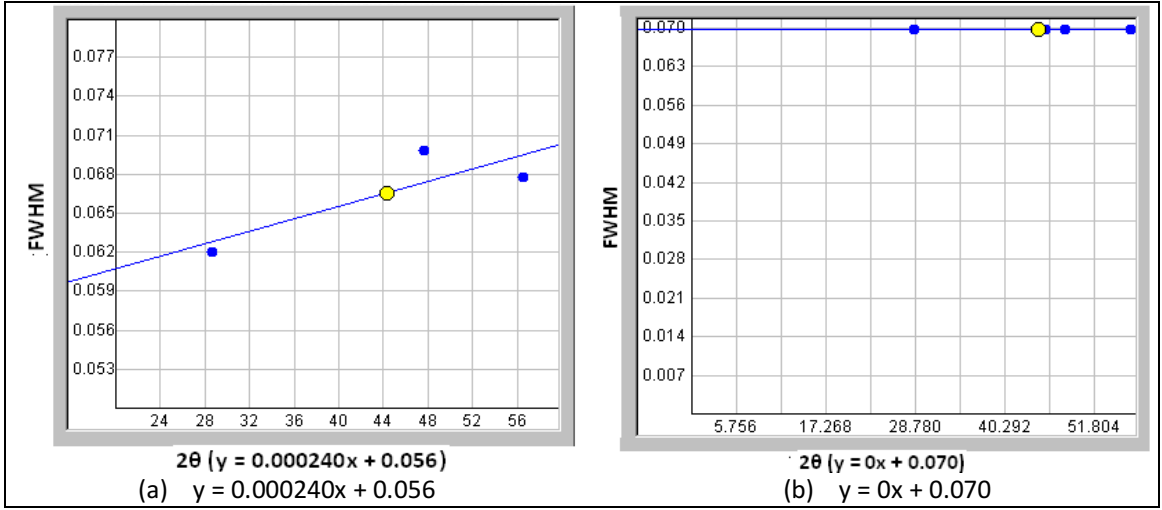

**Figure 2.** Typical Instrumental Broadening. (a) undoped ZnS  (b) 0.1M FeCl$_3$ doped ZnS

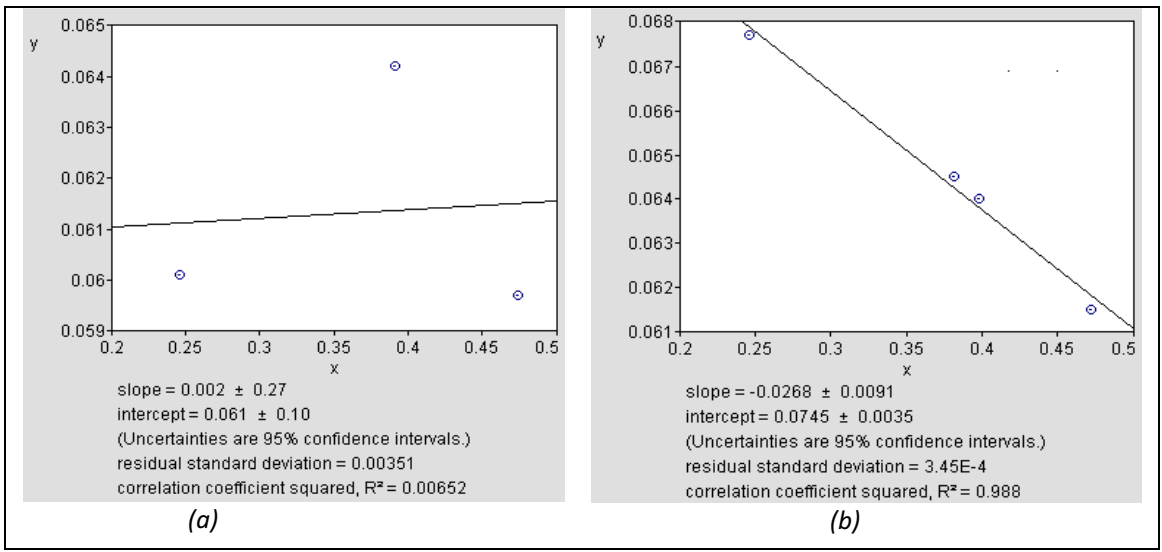

**Figure 3.** W.H.Plot line broadening value.  (a) undoped ZnS  (b) 0.1M FeCl$_3$ doped ZnS

**Table 2.** Data for W.H.Plot of undoped ZnS and 0.1M FeCl$_3$ doped ZnS

| Undoped ZnS | | | 0.1M FeCl$_3$ doped ZnS | | |
|---|---|---|---|---|---|
| θ | sin θ | β cos θ | θ | sin θ | β cos θ |
| 14.242 | 0.246 | 0.0601 | 14.2428 | 0.246 | 0.0677 |
| 23.0546 | 0.3916 | 0.0642 | 22.458 | 0.382 | 0.0645 |
| 28.282 | 0.4738 | 0.05971 | 23.4797 | 0.3984 | 0.064 |
| | | | 28.2058 | 0.4726 | 0.0615 |
| y intercept=0.061 ± 0.10   size=2.27 nm slope=0.002±0.27              strain=0.0005 | | | y intercept=0.0745 ± 0.0035   size=1.86nm slope= 0.0268±0.0091           strain=0.0067 | | |

Line broadening analysis is most accurate when the broadening due to particle size effects is at least twice the contribution due to instrumental broadening. The size range is calculated over which this technique will be most accurate. A rough upper limit is estimated for reasonable accuracy by looking at the particle size that would lead to broadening equal to the instrumental broadening. For example, for Monochromatic Lab X-ray (Cu Kα FWHM ~ 0.05° at 20° 2θ), the accurate Size Range is < 90 nm (900 Å) and the rough Upper Limit is = < 180 nm (1800 Å).

*XRD - Specific Surface Area*

Specific surface area (SSA) is a material property. It is a derived scientific value that can be used to determine the type and properties of a material. It has a particular importance in case of adsorption, heterogeneous catalysis and reactions on surfaces. SSA is the Surface Area (SA) per mass.

$$SSA = \frac{SA_{part}}{V_{part}*density} \quad \text{...................................} (5)$$

Here Vpart is particle volume and SApart is particle SA[12].

$$S = 6*10^3/D_p\rho \quad \text{.........................................} (6)$$

Where S is specific surface area, Dp is size of particles, and ρ is density of particle[13]. ZnS density is 4.09 g cm$^{-3}$. Mathematically, SSA can be calculated using these formulas 5 and 6. Both of these formulas yield same result. Calculated value of prepared ZnS nanoparticles are presented in Table.3.

**Table 3.** Specific Surface Area of Nickel oxide Nanoparticles

| Sample | Particle Size (nm) | Surface Area (nm$^2$) | Volume (nm$^3$) | SSA (m$^2$g$^{-1}$) | SA to Volume Ratio |
|---|---|---|---|---|---|
| undoped ZnS | 2.27 | 16.18 | 6.12 | 646 | 2.643 |
| 0.01M FeCl$_3$ doped ZnS | 2.33 | 17.05 | 6.62 | 630 | 2.575 |
| 0.1M FeCl$_3$ doped ZnS | 1.86 | 10.86 | 3.36 | 788 | 3.232 |

*SEM Images Analysis*

SEM images of Pure ZnS and FeCl$_3$ doped ZnS nanoparticles 10000 magnifications are shown in Figure 4. SEM images show that the prepared nanoparticles were found to be in cluster form. The surface morphology of undoped and FeCl$_3$ doped ZnS nanoparticles, surface morphology of the particles are Cauliflower like structure[14]. In some places, various sizes of the particles (small and large size) are observed[15].

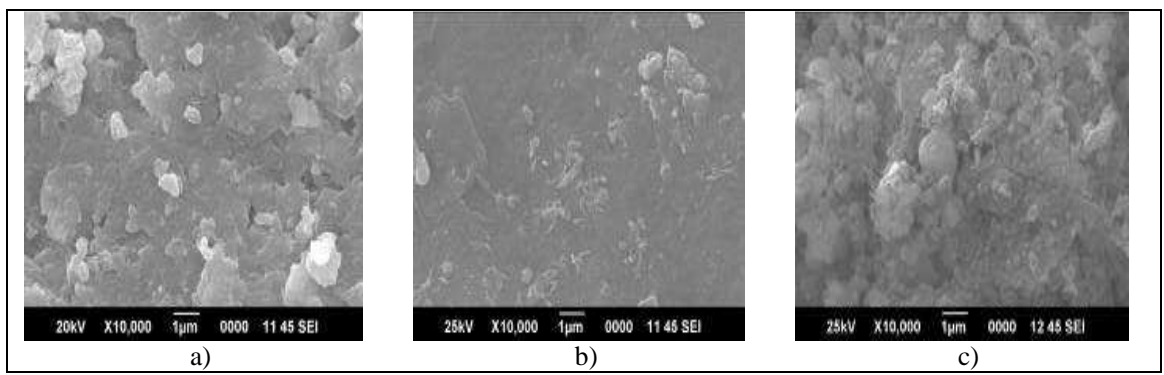

**Figure 4**. SEM images a) undoped ZnS  b) 0.01M FeCl$_3$ doped ZnS c) 0.1M FeCl$_3$ doped ZnS

*UV-Vis measurement*

The UV-Visible absorption spectroscopy of the prepared samples was recorded from the SHIMADZU UV-Visible absorption Spectrometer. Figure 5.a., 5.b. and 5.c. shows the UV-Visible absorption spectrum of undoped and $FeCl_3$ doped ZnS respectively. The absorbance versus wavelength traces for all the samples have been recorded in the range 200-400 nm. The absorption, which corresponds to electron excitation from the valence band to conduction band, can be used to determine the nature and value of the optical band gap. The relation between the absorption coefficient (α) and the incident photon energy (hγ) can be written as[16]

$$(\alpha h\gamma) = A(h\gamma - E_g)^n \qquad (7)$$

Where A is a constant and $E_g$ is the band gap of the material and exponent n depends on the type of transition. For direct allowed n = ½, indirect allowed transition n=2 and for direct forbidden n=3/2.

The absorption edge shift towards the lower value of wavelength (higher energy) while increase in concentration of $FeCl_3$. It is clear that the band gap increases with doping concentration[17]. The trend of observed band gap variation after doping in present study is similar to that reported earlier[16]. In our systems, it is postulated that the rate of nucleation is increased with $FeCl_3$ doping concentration, which produces relatively small particle leading to quantum confinement effect. The particle size of the prepared samples can be determined by the equation,

$$\Delta E_g = E_g^{nano} - E_g^{bulk} = \frac{h^2\pi^2}{2\,MR^2} \qquad (8)$$

Where R is the radius of the particles and M is the effective mass of the system. The particle size of the prepared samples is 5.35 nm, 4.98 nm and 4.13 nm undoped ZnS, 0.01M $FeCl_3$ doped ZnS and 0.1M $FeCl_3$ doped ZnS respectively.

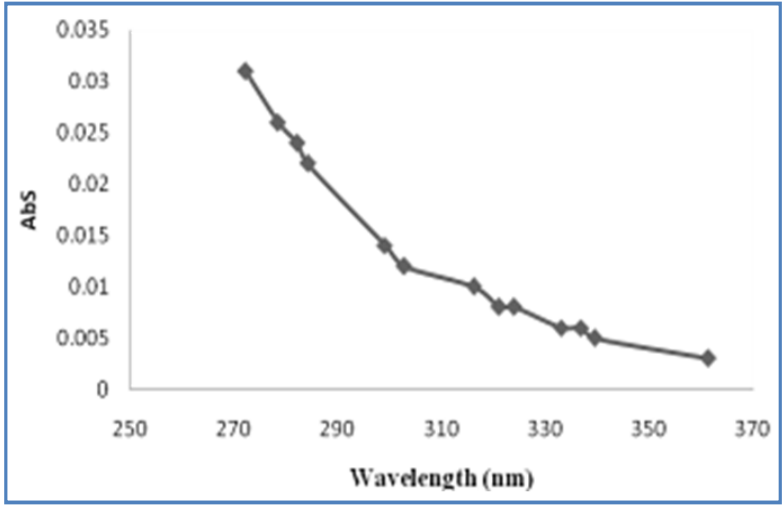

**Figure 5.a.** UV-Vis Spectrum of Undoped ZnS

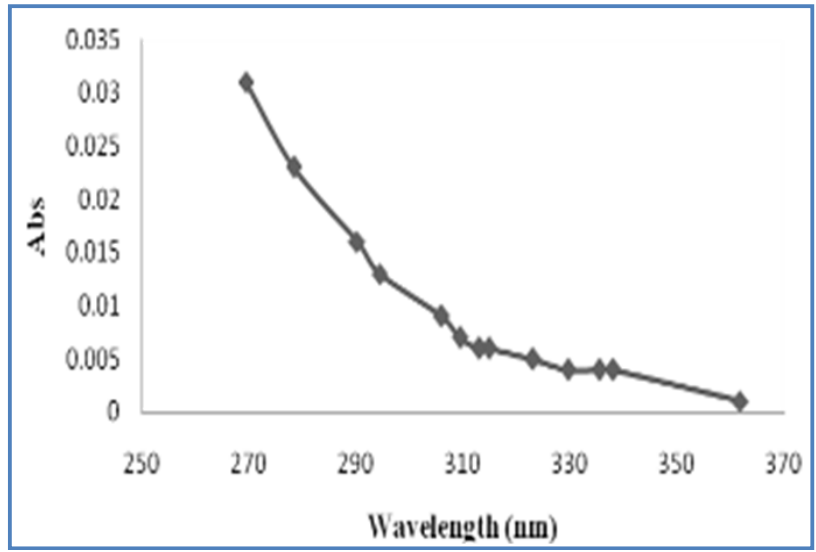

**Figure 5.b.** UV-Vis Spectrum of 0.01M FeCl$_3$ doped ZnS

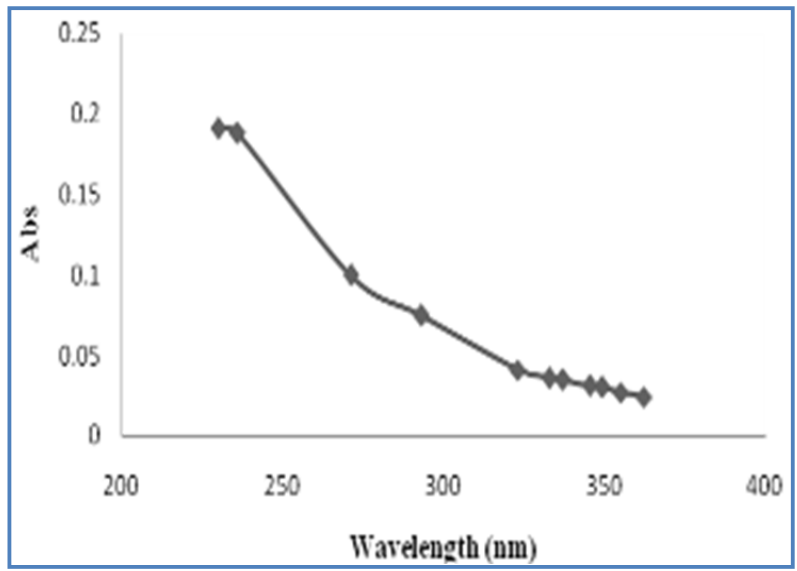

**Figure 5.c.** UV-Vis Spectrum of 0.1M FeCl$_3$ doped ZnS

**Table 4.** Particle size and Energy gap value for Undoped ZnS and FeCl$_3$ doped ZnS

| Sample | Particle size (nm) | Energy gap Value of Samples (eV) | Energy gap Value of Bulk ZnS (cubic) |
|---|---|---|---|
| Undoped ZnS | 5.35 | 4.20 | 3.54 eV |
| 0.01M FeCl$_3$ doped ZnS | 4.98 | 4.28 | |
| 0.1M FeCl$_3$ doped ZnS | 4.13 | 4.55 | |

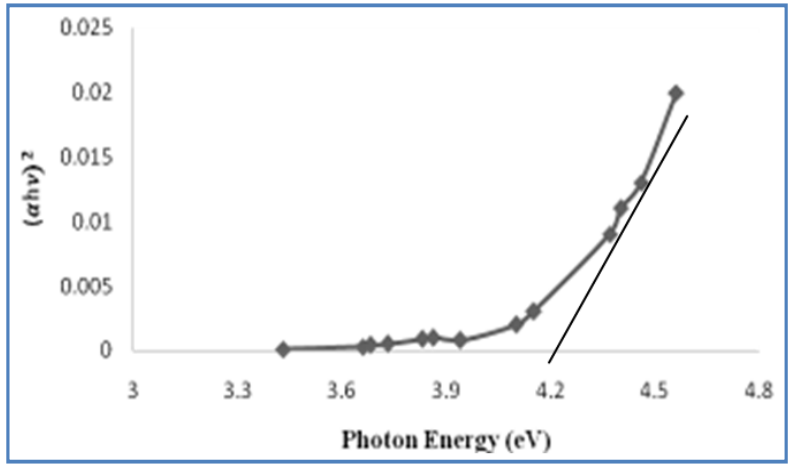

**Figure 6.a.** Energy Band gap Graph of Undoped ZnS

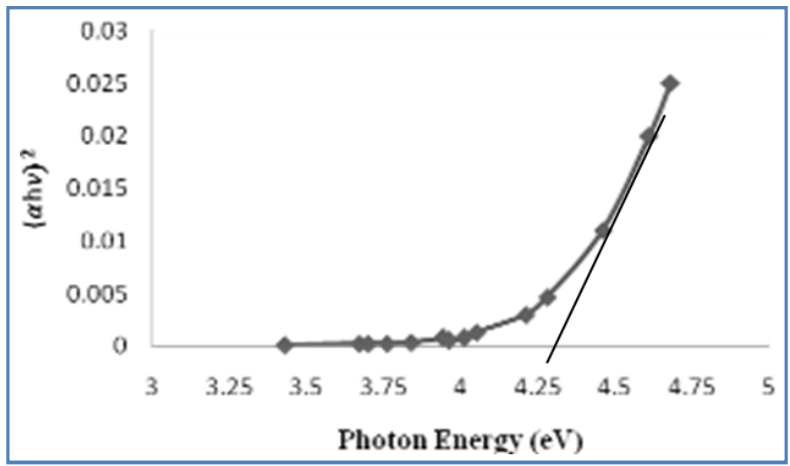

**Figure 6.b**. Energy Band gap Graph of 0.01M $FeCl_3$ doped ZnS

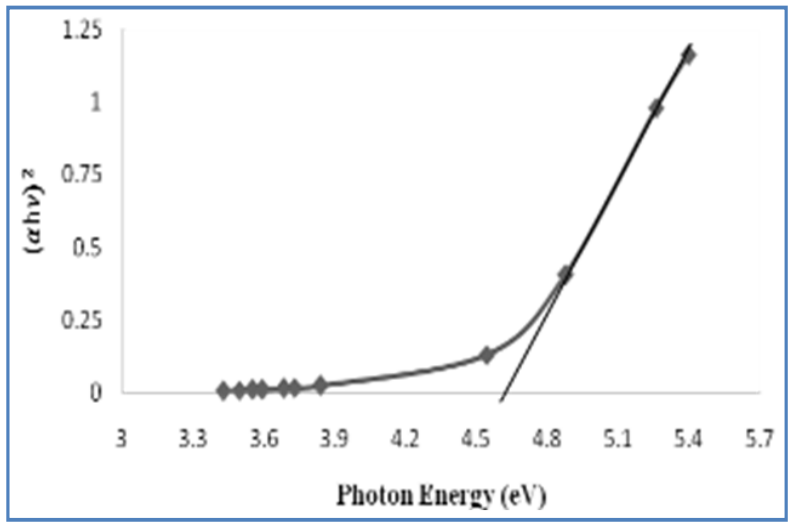

**Figure 6.c.** Energy Band gap Graph of 0.1M $FeCl_3$ doped ZnS

To measure the energy band gap value from the absorption spectra of a graph $(\alpha h\gamma)^2$ versus $(h\gamma)$ is plotted in Figure 6.a., 6.b. and 6.c. for Undoped ZnS, 0.01M and 0.1M FeCl$_3$ doped ZnS nanoparticles. The exact value of the band gap is determined by extrapolating the straight line portion of $(\alpha h\gamma)^2$ vs. $(h\gamma)$ graph to the $h\gamma$ axis. The direct allowed band gap values of ZnS, 0.01M FeCl$_3$ doped ZnS and 0.1 M FeCl$_3$ doped ZnS have been found to be 4.20 eV, 4.28 eV and 4.55 eV respectively. It is noticed that the band gap value of samples are higher than bulk ZnS. The particle size and Energy gap value of Undoped ZnS and FeCl$_3$ doped ZnS samples are given in Table 4.

*Thermal analysis measurement*

Thermal analysis Measurement of prepared nanoparticle is carried out by using Thermo Gravimetric Analysis (TGA), Differential Thermal Analysis (DTA) and Differential Scanning Calorimetry method. TGA/DTA Thermograms and DSC Curve of Pure ZnS and FeCl$_3$ doped ZnS nanoparticles are shown in Figure 7.a. - 7.f. The thermal analysis data of the samples are given in Table 5.

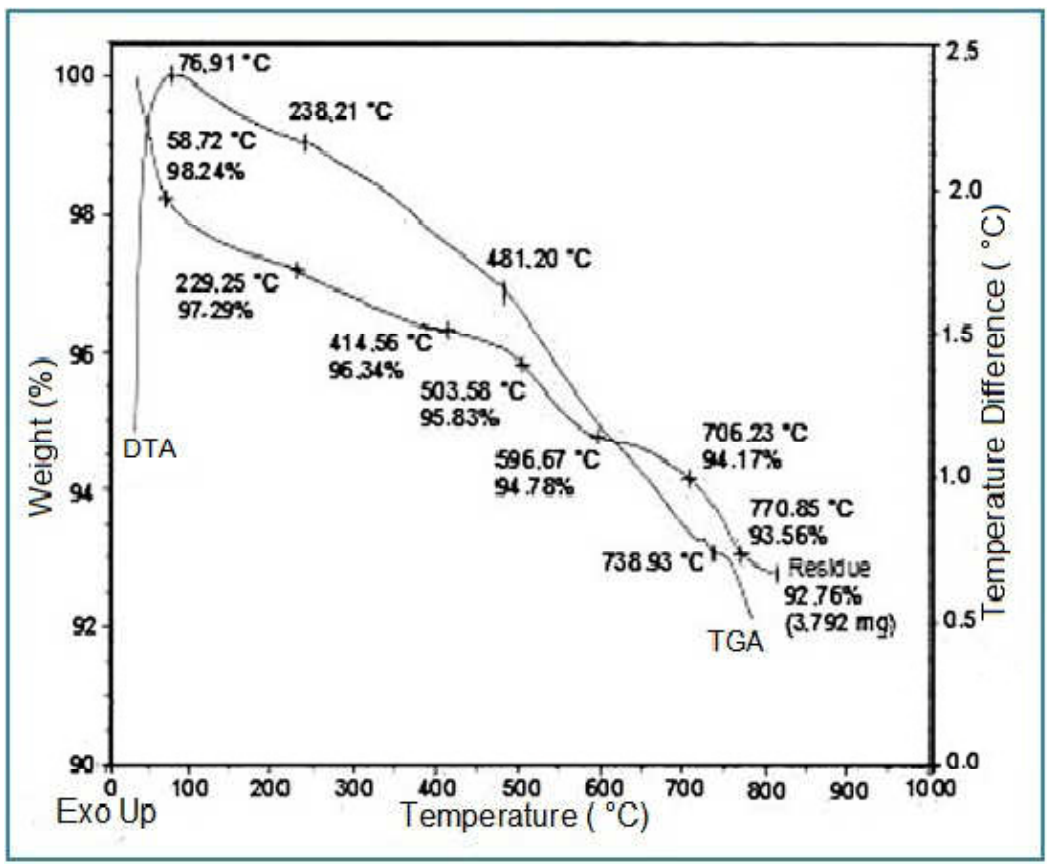

**Figure 7.a.** TGA and DTA Thermograms of Undoped ZnS

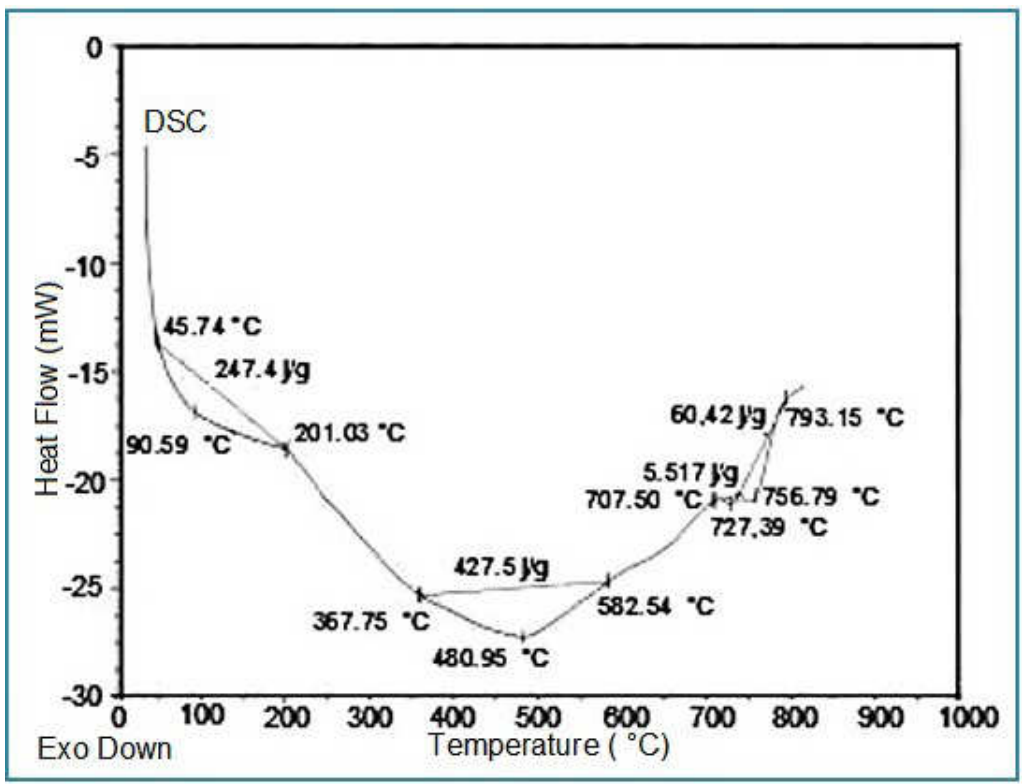

**Figure 7.b.** DSC Curve of Pure ZnS

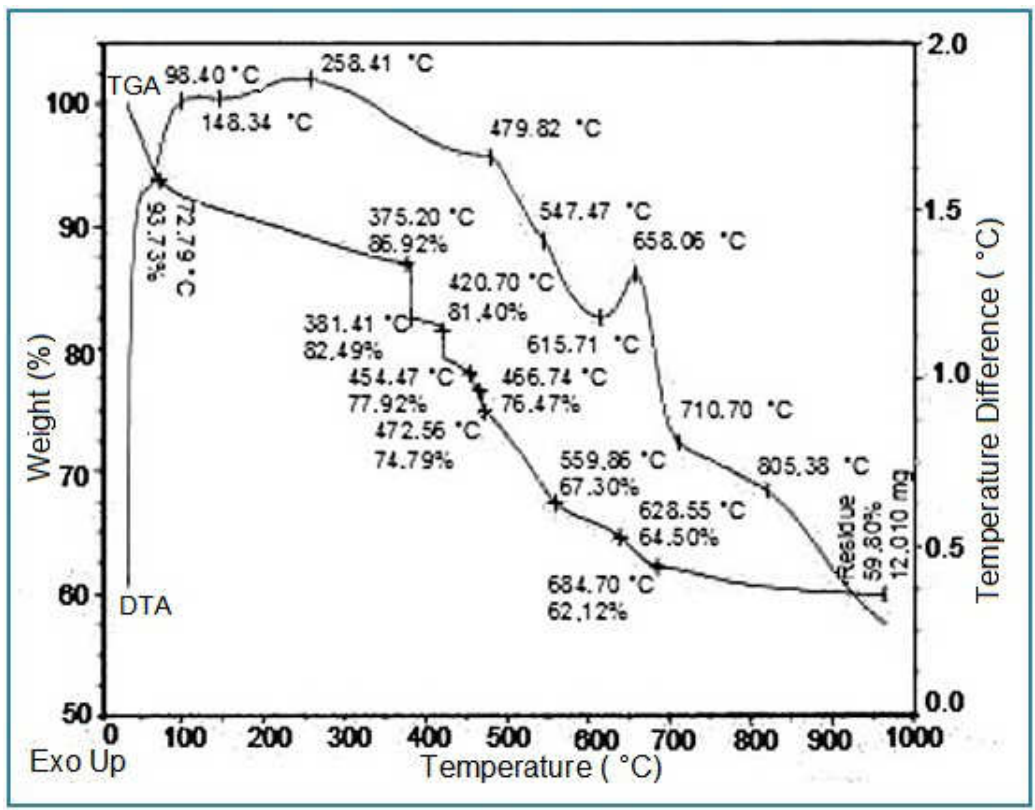

**Figure 7.c.** TGA and DTA Thermograms of 0.01M FeCl$_3$ doped ZnS

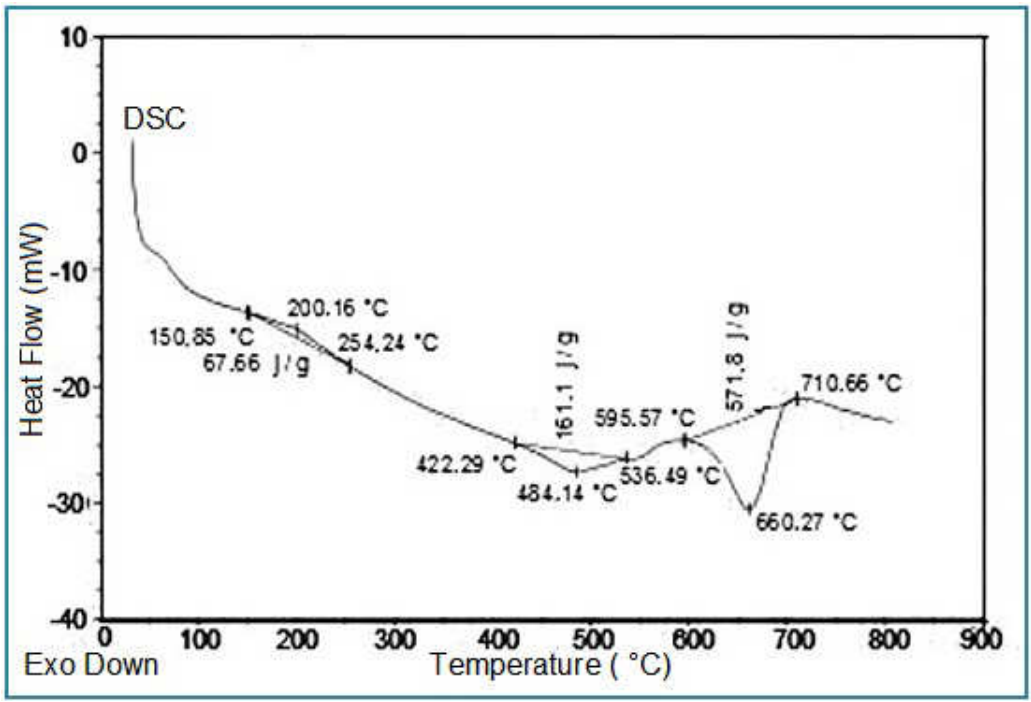

**Figure 7.d.** DSC Curve of 0.01M $FeCl_3$ doped ZnS

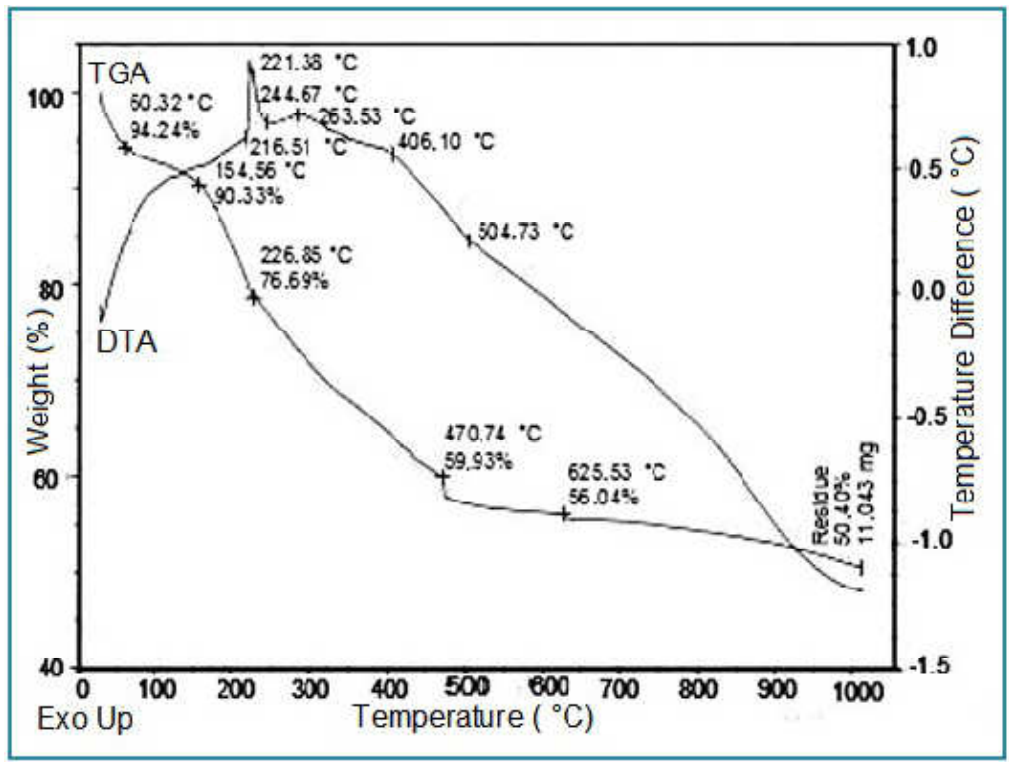

**Figure 7.e.** TGA and DTA Thermograms of 0.1M $FeCl_3$ doped ZnS

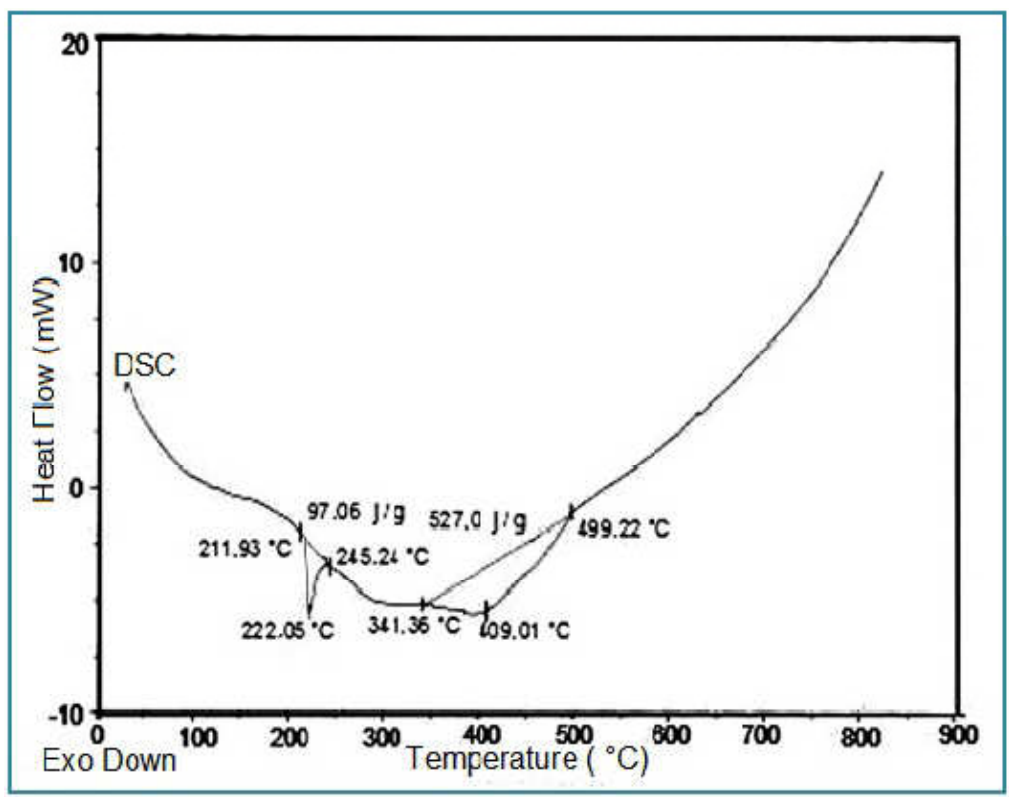

**Figure 7.f.** DSC Curve of 0.1M FeCl$_3$ doped ZnS

**Table 5.** Thermal Analysis Data

| Material | TGA temp (°C) | DSC Peak temp (°C) | DTA Weight loss (%) | Total Weight loss (%) (50°C to 750°C) |
|---|---|---|---|---|
| Undoped ZnS | 481.20 | 480.95 | 2.46 | 7 |
|  | 738.93 | 727.39 | 2.5 |  |
| 0.01M FeCl$_3$ doped ZnS | 258.41 | 200.16 | 8 | 46 |
|  | 479.82 | 484.14 | 27 |  |
|  | 658.06 | 660.97 | 11 |  |
| 0.1M FeCl$_3$ doped ZnS | 221.38 | 222.05 | 24 | 48.5 |
|  | 406.10 | 409.01 | 24.5 |  |

In undoped ZnS and FeCl$_3$ doped ZnS, Weight loss occurs from 50°C to 150°C is due to some water molecule present in the Samples. In Pure ZnS the decomposition occurs at 481.20°C, ZnS is converted into ZnSO$_3$ with weight loss of 2.46% and at 738.93°C ZnSO$_3$ is converted into ZnO with 2.5% weight loss[18]. Therefore, the total weight loss in undoped ZnS is 7% from 50°C to 750°C.

In 0.01M FeCl$_3$ doped ZnS, the decomposition proceeds in three steps. In first step, there is a Endothermic Peak is observed at 200.16°C with weight loss of 8% and also the Second and third step, there is a Exothermic Peak is observed at 484.14°C & 660.97°C with weight loss of 27% & 11% respectively. In 0.01M FeCl$_3$ doped ZnS, the total weight loss occurs of 46% from 50°C to 750°C.

In 0.1 M FeCl$_3$ doped ZnS, the decomposition proceeds in two steps, there is an exothermic peak is observed at 222.05°C & 409.01°C with weight loss of 24% & 24.5% respectively. The total weight loss occurs of 48.5% from 50°C to 750°C.

From thermal analysis, we conclude that the percentage of weight loss will lesser in undoped ZnS compared to FeCl$_3$ doped ZnS. The lesser in weight loss gives the better thermal stability[19]. In FeCl$_3$ doped ZnS, iron will easily oxidized to air in high temperatures. So, undoped ZnS is more thermal stable compared to FeCl$_3$ doped ZnS and also the thermal stability of undoped ZnS is decreased due to increase in FeCl$_3$ concentration.

**Conclusions**

Nanoparticles of ZnS and FeCl$_3$ doped ZnS have been synthesized successfully by simple chemical precipitation method. The XRD of the samples confirm, the formation of Nanoparticles of ZnS and FeCl$_3$ doped ZnS and their cubic structure. Particle size and specific surface area have been calculated from XRD analyses which confirm the nano structure of the samples. The SEM images show that the cauliflower likes structure. Absorption spectra have been obtained using UV-Visible spectrophotometer to find the optical band gap value. Particle size calculated from UV-Vis analysis which is also indicating the nano size of the samples. Thermal analysis measurement of the prepared sample shows pure ZnS is more thermal stable compared to Ferric Chloride doped ZnS.

**Acknowledgement**

The authors express immense thanks to staff & management of *Ayya Nadar Janaki Ammal College*, Sivakasi, India for their valuable suggestions and assistances.